\documentclass[letterpaper, 10 pt, conference]{ieeeconf}

\IEEEoverridecommandlockouts                              

\usepackage{graphics} 
\usepackage{subcaption}  
\usepackage{epsfig} 
\usepackage{amsmath} 
\usepackage{amssymb}  
\usepackage{xcolor}
\usepackage{mathtools}

\newtheorem{rmk}{Remark}

\title{\LARGE \bf
The asymptotic variance of the Hermite-Domain Predictor-Based Subspace Identification method
}

\author{Enrico Barbiero$^{1}$ and Marco Lovera$^{1}$
\thanks{$^{1}$Dipartimento di Scienze e Tecnologie Aerospaziali,
	Politecnico di Milano, Via La Masa 34, 20156 Milano, Italy \;\;\; \quad
    \quad 
     {\tt\small $\{$enrico.barbiero, marco.lovera$\}$@polimi.it}}
}

\begin{document}

\maketitle
\thispagestyle{empty}
\pagestyle{empty}

\begingroup
\renewcommand{\thefootnote}{}
\footnotetext[0]{%
\copyright~2026 IEEE. Personal use of this material is permitted.  Permission from IEEE must be obtained for all other uses, in any current or future media, including reprinting/republishing this material for advertising or promotional purposes, creating new collective works, for resale or redistribution to servers or lists, or reuse of any copyrighted component of this work in other works.%
}
\endgroup


\begin{abstract}
This paper derives analytical expressions for the asymptotic variance of the Hermite-Domain Predictor-Based Subspace Identification (HD-PBSID) method. Unlike continuous-time subspace approaches that rely on intermediate-domain transformations, HD-PBSID identifies continuous-time state-space matrices directly. The derivation first establishes the uncertainty of the estimated state-space matrices and then exploits these results to obtain asymptotic variance expressions. These expressions provide analytical uncertainty bounds for system eigenvalues, which are invariant under similarity transformations and therefore uniquely characterize the identified dynamics. The proposed formulas are first validated in an ideal Hermite-domain setting, where the conditions underlying the derivation are satisfied, and then assessed on noisy time-domain data through Monte Carlo simulations. The predicted bounds show very good
agreement with the eigenvalue distributions in the ideal setting and acceptable agreement in the time-domain case when the innovation sequence is enforced to be white.
\end{abstract}


\section{INTRODUCTION}
System identification aims at deriving models of system dynamics from experimental input-output data. Such models are essential both for reproducing the behavior of the underlying physical system and for designing controllers that ensure stability and performance in practical operation. Since experimental data are generally noisy, identified models are inevitably affected by uncertainty. Quantifying this uncertainty is therefore of central importance, especially in robust control, where performance and stability must be guaranteed despite model mismatch \cite{Doyle1996robustcontrol}.

For subspace identification methods, uncertainty quantification is particularly challenging because the identified state-space matrices are only determined up to a similarity transformation. However, system invariants such as poles and transfer functions are unaffected by similarity transformations and are therefore uniquely identified \cite{overviewSubspace}. Among the most widely used Discrete-Time (DT) subspace methods, which provide DT
Linear Time-Invariant (LTI) models, are MOESP \cite{VerhaegenMOESP}, N4SID \cite{Overschee1996SubspaceIF}, SSARX \cite{JANSSON2005SSARX}, and PBSID \cite{Chiuso2007PBSID}. The latter is especially attractive in practice because it can handle data collected in closed loop. For PBSID, analytical expressions for the asymptotic variance were derived in \cite{CHIUSOASYMPTOTICVARIANCE2006} and later adapted to PBSID$_{opt}$ in \cite{VANWINGERDEN2012ASYMPTOTICVARIANCE}, yielding manageable formulas for the uncertainty of the identified state-space matrices and, consequently, of the system invariants.

Despite their effectiveness, DT subspace methods may be problematic when a Continuous-Time (CT) model is ultimately sought. Indeed, discretization is non-bijective, so the transformation between the \textit{s}-domain and the \textit{z}-domain is not unique \cite{farahBOUNDED_ERROR_CTMODELS}. This motivates the development of methods that identify CT state-space models directly, such as CT-MOESP (\cite{HAVERKAMP_MOESP}, \cite{ohta2004continuous}) and CT-PBSID \cite{Bergamasco2011}. In CT-PBSID, the CT input-output signals are transformed into discrete coefficients through Laguerre filters or Laguerre projections, and the model is first identified in the Laguerre domain before being mapped back to the CT domain through bijective transformations.
Although statistical uncertainty assessment methods such as bootstrap procedures have been proposed for CT-PBSID \cite{bootstrapBergamasco}, deriving analytical asymptotic variance expressions remains difficult. The reason is that CT-PBSID repeatedly time-shifts the input-output signals and projects them onto Laguerre basis functions, which introduces correlations in the innovation sequence \cite{Bergamasco2011}. As a result, the white-noise property underlying the asymptotic variance derivation in \cite{VANWINGERDEN2012ASYMPTOTICVARIANCE} is no longer valid.

The recently developed Hermite-Domain (HD) PBSID \cite{rebollo2026continuoustimepredictorbasedsubspaceidentification} overcomes these limitations. HD-PBSID projects the CT signals onto orthonormal Hermite basis functions without the need of repeatedly shifting the signals and exploits the derivative recurrence relations of Hermite functions to identify the CT state-space matrices directly, without any intermediate-domain realization. Since no repeated time shifting is involved, the correlations that affect CT-PBSID innovations are avoided, at least from the theoretical standpoint, and the white-noise innovation property is preserved. This opens the way to the direct derivation of asymptotic variance expressions for the CT matrices identified by HD-PBSID.

In this paper, analytical expressions for the asymptotic variance of the HD-PBSID method are derived by extending the results of \cite{VANWINGERDEN2012ASYMPTOTICVARIANCE}. The proposed formulas are first validated in an ideal setting, where possible sources of mismatch with the theoretical conditions are excluded, namely identification bias, projection-induced errors, and numerically induced coloring of the innovation sequence. To this end, simulations are carried out directly in the Hermite domain, and the true CT state-space matrices are inserted into the asymptotic variance formulas, which are then converted into uncertainty bounds for the system eigenvalues. Further validation is then performed on time-domain simulations, where the above deviations from the theoretical conditions are naturally present. In this more realistic setting, particular attention is devoted to the role of the innovation sequence: besides using the residual-based innovation directly, a second validation is carried out by enforcing a white-noise innovation with the same estimated covariance, showing that this treatment significantly improves the agreement between the analytical predictions and the Monte Carlo (MC) results.
This paper is organized as follows. Section \ref{sec2: problem statement} states the problem, Section \ref{sec3: Hermite-Domain Predictor Based Subspace Identification} summarizes the HD-PBSID method, Section \ref{sec4: Asymptotic variance derivation} derives the uncertainty of the estimated state-space matrices and the corresponding asymptotic variance expressions, Section \ref{sec5: Asymptotic variance validation in the Hermite domain} validates the formulas in the Hermite domain, Section \ref{sec6: Test example} presents the time-domain validation, and Section \ref{sec7: Conclusions} concludes the paper.

\section{Problem statement}
\label{sec2: problem statement}
Consider the LTI stochastic CT system in innovation form:
\begin{equation}
\label{eq:LTI stochastic system in innovation form}
\begin{split}
d\hat{x}(t) &= A\hat{x}(t)\,dt + Bu(t)\,dt + K\,de(t), 
\qquad x(0) = x_0, \\
d\hat{z}(t) &= C\hat{x}(t)\,dt + Du(t)\,dt, \\
y(t)\,dt &= d\hat{z}(t) + de(t),
\end{split}
\end{equation}
with $\hat{x}(t) \in \mathbb{R}^{n_x}$, $u(t) \in \mathbb{R}^{n_u}$, and $y(t) \in \mathbb{R}^{n_y}$, and innovation $e(t) \in \mathbb{R}^{n_y}$ is a Wiener process.
Let $\underbar{f}=[\underbar{f}_1^\top,$ $\dots,$ $\underbar{f}_{n_f}^\top]^\top \in \mathbb{R}^{n_f \times n_{c}}$, where $n_{c} = n_{max} + 1$, denote the projections of a generic function $f(t):\mathbb{R}\to\mathbb{R}^{n_f}$ defined on $\mathbb{T} = [-\mathcal{T},\mathcal{T}]$ onto the finite set of orthonormal Hermite basis functions $\{h_n\}_{n=0}^{n_{max}}$. Each row in $\underbar{f}$ displays the projection coefficients of the corresponding channel of $f(t)$: for example, $\underbar{f}_1$ contains the projections of the first channel of $f(t)$. Each coefficient $f_n \in \mathbb{R}^{n_f}$ is given by $f_n = \int_{\mathbb{T}} f(t) h_n(t) \, dt$. Hence, $f(t)$ can be expressed as $f(t) \simeq \underbar{f}\boldsymbol{h}(t)$, where $\boldsymbol{h}(t) = [h_{0}(t), \dots, h_{n_{max}}(t)]^\top$. 
Given that the derivative of the $n$-th normalized Hermite function depends on the previous and on the subsequent Hermite functions as $d(h_n(t))/dt =  (1/\sqrt{2}) \left(\sqrt{n} h_{n-1}(t) - \sqrt{n+1} h_{n+1}(t)\right)$ (see \cite{szego1959orthogonal}), then the derivative of the generic function $f(t)$ is $d(f(t))/dt \simeq \sum_{n=0}^{n_{max}}{f_n ( \frac{1}{\sqrt{2}} (\sqrt{n} h_{n-1}(t) - \sqrt{n+1} h_{n+1}(t)))} = \underbar{f}\mathcal{D}\boldsymbol{h}(t)$. The derivative operator $\mathcal{D}\in\mathbb{R}^{n_c\times n_c}$ is induced by the Hermite recurrence relations, and its entries can be compactly written as $(\mathcal{D})_{ij} = \sqrt{(i-1)/2}\,\delta_{j,i-1}-\sqrt{i/2}\,\delta_{j,i+1}$, with $ i,j=1,\dots,n_c$ and $\delta$ being the Kronecker delta. 
Projecting the CT signals $\hat{x}(t)$, $u(t)$, $y(t)$, and $de(t)$ of (\ref{eq:LTI stochastic system in innovation form}) onto $\{h_n\}_{n=0}^{n_{max}}$ and using the above Hermite derivative recurrence relations yields the Hermite-coefficient LTI system \cite{rebollo2026continuoustimepredictorbasedsubspaceidentification}:
\begin{equation}
\label{eq: CT LTI dynamics with Hermite expansions}
\begin{split}
\underbar{x} \mathcal{\tilde{D}} &= A \underbar{x} + B \underbar{u} + K\underbar{e} \\
\underbar{y} &= C \underbar{x} + D \underbar{u} + \underbar{e},
\end{split} 
\end{equation}
where \(\boldsymbol{h}(t)\) does not appear because the equations are written after projection, directly in terms of the Hermite coefficients 
$\underbar{x}\in\mathbb{R}^{n_x\times n_c}$, $\underbar{y}\in\mathbb{R}^{n_y\times n_c}$, $\underbar{u}\in\mathbb{R}^{n_u\times n_c}$, $\underbar{e}\in\mathbb{R}^{n_y\times n_c}$, each defined as the above generic $\underbar{f}$. The Hermite-domain innovation sequence $\underbar{e} = [\underbar{e}_1^\top,\dots,\underbar{e}_{n_y}^\top]^\top\in\mathbb{R}^{n_y\times n_c}$ is a zero-mean white Gaussian process, with covariance matrix $\Lambda\in\mathbb{R}^{n_y\times n_y}$. This property is shown in the Appendix.
Matrices $A$, $B$, $C$, and $D$ are the same as in (\ref{eq:LTI stochastic system in innovation form}), which means they are preserved in CT form, differently from typical CT subspace methods where the matrices are transformed into the domain of the orthonormal basis. The derivative operator in (\ref{eq: CT LTI dynamics with Hermite expansions}) is defined as $\mathcal{\tilde{D}}=\mathcal{D}/\alpha$, with $\alpha = t_u/(2\sqrt{2n_{max}+1})$, where $t_u$ denotes the time duration of the signals. The scaling parameter $\alpha$ is introduced to account for the effective support of $\{h_n\}_{n=0}^{n_{max}}$.
The state equation can then be rewritten as $\underbar{x} \mathcal{D} = \bar{A} \underbar{x} + \bar{B} \underbar{z},$
where $\bar{A} = A-KC$, $\bar{B} =  [B - KD \quad K]$ and $\underbar{z} = [\underbar{u}^\top \quad \underbar{y}^\top]^\top$. 
To make the system suitable for PBSID, parameters $\beta$ and $\gamma$
are introduced to modify $\bar{A}$ so that its eigenvalues lie inside the
unit circle, as discussed in
\cite{rebollo2026continuoustimepredictorbasedsubspaceidentification}.
Defining $\mathcal{D}' = (\mathcal{D}/{\alpha} + \beta I)/\gamma$, the
modified state-space system becomes:
\begin{equation}
\begin{split}
\label{eq: LTI dynamics modified for stability}
\underbar{x} \mathcal{D}' &= \frac{\bar{A} + \beta I}{\gamma} \underbar{x} + \frac{\bar{B}}{\gamma} \underbar{z} = \bar{A}' \underbar{x} + \bar{B}' \underbar{z}
\\
\underbar{y} &= C \underbar{x} + D \underbar{u} + \underbar{e},
\end{split}
\end{equation}
which can be processed by the HD-PBSID algorithm, summarized in Section~\ref{sec3: Hermite-Domain Predictor Based Subspace Identification}.
Given $u(t)$ and $y(t)$, HD-PBSID identifies $A$, $B$, $C$, $D$, and $K$ up to a similarity transformation $T \in \mathbb{R}^{n_x \times n_x} : T^{-1}AT \; , \; T^{-1}B \; , \; T^{-1}K \; , \; CT \; , \; D$.
\\Let $\widehat{A}$, $\widehat{B}$, $\widehat{K}$, $\widehat{C}$, and $\widehat{D}$ denote the CT matrices identified by HD-PBSID, and define the uncertainty of a generic estimate as $\Delta\widehat{{(\; \cdot \; )}} = \widehat{(\; \cdot \; )} - (\; \cdot \; )$, where $(\cdot)$ denotes the corresponding true quantity up to a similarity transformation. The problem is then to determine the asymptotic variance:
\begin{equation}
    \operatorname{plim} \Delta\widehat{\Theta}\Delta\widehat{\Theta}^\top,
\end{equation}
where $\operatorname{plim}$ stands for `probability limit' and:
\begin{equation}
\label{eq: uncertainty of the estimates}
\begin{split}
    \Delta\widehat{\Theta} = [ \operatorname{vec}(\widehat{A} - T^{-1}AT)^\top, \operatorname{vec}(\widehat{B} - T^{-1}B)^\top, \\
    \operatorname{vec}(\widehat{K} - T^{-1}K)^\top, \operatorname{vec}(\widehat{C} - CT)^\top, \operatorname{vec}(\widehat{D} - D)^\top]^\top .
\end{split}
\end{equation}

\section{Hermite-Domain Predictor Based Subspace Identification}
\label{sec3: Hermite-Domain Predictor Based Subspace Identification}
The main steps of HD-PBSID are briefly summarized below. For further details, see \cite{rebollo2026continuoustimepredictorbasedsubspaceidentification}.
\\Differentiating $p$ times the state equation in (\ref{eq: LTI dynamics modified for stability}), one gets:
\begin{equation}
\label{eq: state equation with multiple differentiation}
\underbar{x} \mathcal{D}'^p = \bar{A}'^p \underbar{x} + \bar{B}' \underbar{z} \mathcal{D}'^{p-1} + \bar{A}' \bar{B}' \underbar{z} \mathcal{D}'^{p-2}  + \cdots +  \bar{A}'^{p-1} \bar{B}' \underbar{z},
\end{equation}
where $p$ is the so-called past window length, chosen so that $\bar{A}'^p \approx 0$. By multiplying equation (\ref{eq: state equation with multiple differentiation}) by $\mathcal{D}'^{-p}$, the state equation reduces to:
\begin{equation}
\label{eq: integral version of state equation}
\underbar{x} \simeq \begin{bmatrix} \bar{A}'^{p-1} \bar{B}' & \cdots &\bar{A}' \bar{B}' & \bar{B}' \end{bmatrix} \begin{bmatrix} \underbar{z} \mathcal{D}'^{-p} \\ \vdots \\ \underbar{z} \mathcal{D}'^{-2}  \\ \underbar{z} \mathcal{D}'^{-1} \end{bmatrix} = \mathcal{K}^p\mathcal{Z},
\end{equation}
where $ \mathcal{K}^p = \begin{bmatrix} \bar{A}'^{p-1} \bar{B}' & \cdots &\bar{A}' \bar{B}' & \bar{B}' \end{bmatrix}$ is the reversed extended controllability matrix. Thus, the output equation becomes:
\begin{equation}
\underbar{y} = C \mathcal{K}^p\mathcal{Z} + D  \underbar{u} + \underbar{e},
\end{equation}
which poses the least-squares problem: 
\begin{equation}
\label{eq:first least squares}
\min_{C \mathcal{K}^p, D} \left\| \underbar{y} - C \mathcal{K}^p \mathcal{Z} - D\underbar{u} \right\|_F,
\end{equation}
with $\left\| \cdots \right\|_F$ representing the Frobenius norm.
The solution is as follows:
\begin{equation}
\label{eq:first least squares}
        [\widehat{C \mathcal{K}^p} \; , \; \widehat{D}] = \underbar{y}\begin{bmatrix} \mathcal{Z} \\ \underbar{u} \end{bmatrix}^{\dagger},
\end{equation}
where $\dagger$ denotes the Moore–Penrose pseudoinverse.
\\Given the definition of the extended observability matrix
$\Gamma^f = \begin{bmatrix} C^\top & (C\bar{A}')^\top & \cdots & (C \bar{A}'^{f-1})^\top \end{bmatrix}^\top$,
with future window length $f$, one obtains:
\begin{equation}
\label{eq:simplified observability times controllability}
\Gamma^f \mathcal{K}^p \simeq 
\begin{bmatrix} 
C \bar{A}'^{p-1}\bar{B} & C \bar{A}'^{p-2}\bar{B}' & \cdots & C \bar{B}' \\
0 & C \bar{A}'^{p-1}\bar{B}' & \cdots & C \bar{A}' \bar{B}' \\
\vdots & & \ddots & \vdots \\
0 & \cdots & & C \bar{A}'^{p-1}\bar{B}'
\end{bmatrix},
\end{equation}
where $\bar{A}'^k \approx 0$ for $k \geq p$ and $f=p$ are used; the matrix entries are obtained directly from $\widehat{C\mathcal{K}^p}$. Multiplying (\ref{eq: integral version of state equation}) by $\Gamma^f$ gives $\Gamma^f\underbar{x} \simeq \Gamma^f\mathcal{K}^p\mathcal{Z}$, whose known right-hand side is then factorized by Singular Value Decomposition (SVD):
\begin{equation}
\label{eq:SVD}
\Gamma^f\mathcal{K}^p \mathcal{Z} = \begin{bmatrix}
\mathcal{U}_n & \mathcal{U}_0 \end{bmatrix} \begin{bmatrix}
\Sigma_n & 0 \\ 0 & \Sigma_0 \end{bmatrix}
\begin{bmatrix}
\mathcal{V}^\top_n \\
\mathcal{V}^\top_0
\end{bmatrix},
\end{equation}
where $\Sigma_n$ contains the $n$ largest singular values, used to estimate the system order, and $\mathcal{V}_n^\top$ is the corresponding row space. The state sequence is then estimated as:
\begin{equation}
\label{eq:state sequence estimate via SVD}
\hat{\underbar{x}} \simeq \Sigma^{\frac{1}{2}}_n\mathcal{V}^\top_n = \Sigma^{-\frac{1}{2}}_n\mathcal{U}^\top_n\Gamma^f\mathcal{K}^p \mathcal{Z},
\end{equation}
which contains a similarity transformation such that $\hat{\underbar{x}} = T\underbar{x} + \Delta\hat{\underbar{x}}$. Matrix $C$ is thus computed from $\min_{C} \|  \underbar{y} - C \hat{\underbar{x}} - \widehat{D} \underbar{u} \|_F $.
The estimate of the innovation sequence $\underbar{e}$ follows:
\begin{equation}
\label{eq:innovation data matrix estimation}
\hat{\underbar{e}} = \underbar{y} - \widehat{C} \hat{\underbar{x}} - \widehat{D} \underbar{u} .
\end{equation}
Solving $\min_{A,B,K} \left\| \hat{\underbar{x}} \mathcal{\tilde{D}} - A \hat{\underbar{x}} - B \underbar{u} - K\hat{\underbar{e}} \right\|_F$,
from the state equation in (\ref{eq: CT LTI dynamics with Hermite expansions}), provides the remaining matrices: 
\begin{equation}
\label{eq:final least squares}
        [\widehat{A} \; , \; \widehat{B} \; , \; \widehat{K}] = \hat{\underbar{x}} \mathcal{\tilde{D}}\begin{bmatrix} \hat{\underbar{x}} \\ \underbar{u} \\ \hat{\underbar{e}} \end{bmatrix}^{\dagger}.
\end{equation}
\begin{rmk}
\label{rmk:remark2}
For the estimates of the state-space matrices to be consistent, under the asymptotic regime $N\to\infty, \; n_c\to\infty$ the following convergences in probability have to hold: 
\begin{equation}
\operatorname*{plim}_{\{N,n_c\} \to \infty}
[\Delta\widehat{A},\Delta\widehat{B},\Delta\widehat{K},\Delta\widehat{C},\Delta\widehat{D}]_{N,n_c}
= 0 ,
\end{equation}
where $N$ is the number of time-domain samples. 
\end{rmk}

\section{Asymptotic variance derivation}
\label{sec4: Asymptotic variance derivation}
The uncertainty of the estimates of the state-space matrices $\Delta\widehat{\Theta}$ (equation (\ref{eq: uncertainty of the estimates})) is first derived; the expression is then exploited to obtain the analytical formulas of the asymptotic variance, by also leveraging the whiteness of the innovation.
\subsection{Uncertainty of the estimates}
To derive $\Delta\widehat{A}, \, \Delta\widehat{B}, \, \Delta\widehat{K}$, the state equation of system (\ref{eq:LTI stochastic system in innovation form}) is considered and the relation $(\; \; ) = \widehat{(\; \; )} - \Delta\widehat{{(\; \; )}}$ is substituted into the unknown terms, yielding the following expansion:
\begin{equation}
    \label{eq:second least squares}
    \begin{split}
    &(\hat{\underbar{x}} - \Delta\hat{\underbar{x}}) \mathcal{\tilde{D}} - (\widehat{A} - \Delta\widehat{A}) (\hat{\underbar{x}} - \Delta\hat{\underbar{x}}) - (\widehat{B} - \Delta\widehat{B}) \underbar{u} + \\& - (\widehat{K} - \Delta\widehat{K}) (\hat{\underbar{e}} - \Delta\hat{\underbar{e}}) \approx \\ & \approx -\Delta\hat{\underbar{x}}\mathcal{\tilde{D}} + \widehat{A} \Delta\hat{\underbar{x}} + \Delta\widehat{A}\hat{\underbar{x}} + \Delta\widehat{B}\underbar{u} + \widehat{K} \Delta\hat{\underbar{e}} + \Delta\widehat{K}\hat{\underbar{e}} \approx 0 ,   
    \end{split}
\end{equation}
where the final expression  follows from $\hat{\underbar{x}}\mathcal{\tilde{D}} -\widehat{A}\hat{\underbar{x}} -\widehat{B}\underbar{u} - \widehat{K}\hat{\underbar{e}} = 0$ and from neglecting the second-order terms $\Delta\widehat{A}\Delta\hat{\underbar{x}}$ and $\Delta\widehat{K}\Delta\hat{\underbar{e}}$. This expression leads to the following least-squares problem:
 \begin{equation}
    \label{eq:second least squares}
    \min_{\Delta{A},\Delta{B},\Delta{K}} \left\| \Delta\hat{\underbar{x}}\mathcal{\tilde{D}} - \widehat{A} \Delta\hat{\underbar{x}} - \Delta\widehat{A}\hat{\underbar{x}} - \Delta\widehat{B}\underbar{u} - \widehat{K} \Delta\hat{\underbar{e}} - \Delta\widehat{K}\hat{\underbar{e}} \right\|_{\mathrm{F}}
\end{equation}
whose solution is the following:
\begin{equation}
\label{eq: uncertainty on state equation}
    [\Delta\widehat{A}, \; \Delta\widehat{B}, \; \Delta\widehat{K}] = [\Delta\hat{\underbar{x}}\mathcal{\tilde{D}} - \widehat{A} \Delta\hat{\underbar{x}} - \widehat{K} \Delta\hat{\underbar{e}}]\begin{bmatrix} \hat{\underbar{x}} \\ \underbar{u} \\ \hat{\underbar{e}}  \end{bmatrix}^{\dagger}.
\end{equation}
Analogously, the expansion can be applied to the output equation of system (\ref{eq:LTI stochastic system in innovation form}):
\begin{equation}
    \begin{split}
    &y - (\widehat{C} - \Delta\widehat{C})(\hat{\underbar{x}} - \Delta\hat{\underbar{x}}) - (\widehat{D} - \Delta\widehat{D})\underbar{u} - \hat{\underbar{e}} + \Delta\hat{\underbar{e}} \approx \\
    & \approx \widehat{C}\Delta\hat{\underbar{x}} + \Delta\widehat{C}\hat{\underbar{x}} + \Delta\widehat{D}\underbar{u} + \Delta\hat{\underbar{e}} \approx 0,
    \end{split}
\end{equation}
where $y - \widehat{C}\hat{\underbar{x}} - \widehat{D}\underbar{u} - \hat{\underbar{e}} = 0$ has been used and the second-order term $\Delta\widehat{C}\Delta\hat{\underbar{x}}$ has been neglected. The corresponding least-squares problem is:
\begin{equation}
    \label{eq:second least squares}
    \min_{\Delta{C},\Delta{D}} \left\| -\widehat{C}\Delta\hat{\underbar{x}} - \Delta\widehat{C}\hat{\underbar{x}} - \Delta\widehat{D}\underbar{u} - \Delta\hat{\underbar{e}} \right\|_F,
\end{equation}
with solution:
\begin{equation}
\label{eq: uncertainty on output equation}
        [\Delta\widehat{C} \; , \; \Delta\widehat{D}] = [-\widehat{C}\Delta\hat{\underbar{x}} - \Delta\hat{\underbar{e}}] \begin{bmatrix} \hat{\underbar{x}} \\ \underbar{u} \end{bmatrix}^{\dagger}.
\end{equation}
Both (\ref{eq: uncertainty on state equation}) and (\ref{eq: uncertainty on output equation}) depend on $\Delta\hat{\underbar{x}}$ and $\Delta\hat{\underbar{e}}$. To derive their expressions, the uncertainty on the estimates of $C \mathcal{K}^p$ and $D$ is first obtained:
\begin{equation}
\label{eq: uncertainty of CK^p and D}
    \begin{split}
        &[\Delta\widehat{C \mathcal{K}^p} \; , \;  \Delta\widehat{D}] = [\widehat{C \mathcal{K}^p} \; , \;  \widehat{D}] - [C \mathcal{K}^p \; , \;  D] = \\ &= \underbar{y}\begin{bmatrix} \mathcal{Z} \\ \underbar{u} \end{bmatrix}^{\dagger} - [C \mathcal{K}^p \; , \;  D] = \left([C \mathcal{K}^p \; , \; D] \begin{bmatrix} \mathcal{Z} \\ \underbar{u} \end{bmatrix} + \underbar{e} \right)\begin{bmatrix} \mathcal{Z} \\ \underbar{u} \end{bmatrix}^{\dagger} + \\ & - [C \mathcal{K}^p \; , \;  D] = \underbar{e}\begin{bmatrix} \mathcal{Z} \\ \underbar{u} \end{bmatrix}^{\dagger} = \underbar{e}\mathcal{Z}_{\mathrm{ext}}^{\dagger} \;.
    \end{split}
\end{equation}
The uncertainty of the estimate of $\underbar{e}$ now follows:
\begin{equation}
\label{eq: uncertainty on e}
    \begin{split}
        \Delta\hat{\underbar{e}} &= \hat{\underbar{e}} - \underbar{e} = -\widehat{C \mathcal{K}^p}\mathcal{Z} +C \mathcal{K}^p\mathcal{Z} - \widehat{D}\underbar{u} + D  \underbar{u} = \\& = - [\Delta \widehat{C \mathcal{K}^p} \; , \; \Delta \widehat{D}] \mathcal{Z}_{\mathrm{ext}} = - \underbar{e}\mathcal{Z}_{\mathrm{ext}}^{\dagger}\mathcal{Z}_{\mathrm{ext}}\;.
    \end{split}
\end{equation}
To obtain the uncertainty of the estimate of $\underbar{x}$, it is of interest to isolate the following expression from equation (\ref{eq: uncertainty of CK^p and D}):
\begin{equation}
\label{eq: expression of delta CK^p}
\Delta\widehat{C \mathcal{K}^p}
=
\left(
\underbar{e}
\mathcal{Z}_{\mathrm{ext}}^{\dagger}
\right)_{1:n_y,\;1:p\cdot(n_u+n_y)}
=
\underbar{e} Z_{\mathrm{ext,part}}^{\dagger}\;.
\end{equation}
By partitioning $Z_{\mathrm{ext,part}}$ into $p$ matrices of $n_u+n_y$ columns each, the following expression can be computed:
\begin{equation}
\label{eq: uncertainty of CK^p with partitioning}
        \Delta\widehat{\Gamma^f \mathcal{K}^p} = \begin{bmatrix}
        \underbar{e}\mathcal{Z}_{\mathrm{ext,part},1}^{\dagger} & \underbar{e}\mathcal{Z}_{\mathrm{ext,part},2}^{\dagger} & \cdots & \underbar{e}\mathcal{Z}_{\mathrm{ext,part},p}^{\dagger} \\
        0 & \underbar{e}\mathcal{Z}_{\mathrm{ext,part},1}^{\dagger} & \cdots & \underbar{e}\mathcal{Z}_{\mathrm{ext,part},p-1}^{\dagger} \\ \vdots & \vdots &  \ddots & \vdots \\
        0 & 0 & \cdots & \underbar{e}\mathcal{Z}_{\mathrm{ext,part},1}^{\dagger}
        \end{bmatrix}
\end{equation}
The uncertainty of the estimate of $\underbar{x}$ is now known, since its expression depends on $\Delta\widehat{\Gamma^f \mathcal{K}^p}$:
\begin{equation}
    \label{eq: delta of state sequence estimate via SVD}
    {\Delta\hat{\underbar{x}}} \simeq \Sigma^{-\frac{1}{2}}_n\mathcal{U}^\top_n\Delta\widehat{\Gamma^f \mathcal{K}^p}\mathcal{Z}.
\end{equation}
Given the following matrix:
\begin{equation}
        \mathcal{Q} \vcentcolon  = \begin{bmatrix}
        \mathcal{Z}_{\mathrm{ext,part},1}^{\dagger} & \mathcal{Z}_{\mathrm{ext,part},2}^{\dagger} & \cdots & \mathcal{Z}_{\mathrm{ext,part},p}^{\dagger} \\
        0 & \mathcal{Z}_{\mathrm{ext,part},1}^{\dagger} & \cdots & \mathcal{Z}_{\mathrm{ext,part},p-1}^{\dagger} \\ \vdots & \vdots &  \ddots & \vdots \\
        0 & 0 & \cdots & \mathcal{Z}_{\mathrm{ext,part},1}^{\dagger}
        \end{bmatrix}\mathcal{Z},
\end{equation}
and letting $\mathcal{S} \vcentcolon  = \Sigma^{-\frac{1}{2}}_n\mathcal{U}^\top_n$, equation (\ref{eq: delta of state sequence estimate via SVD}) can be rewritten as:
\begin{equation}
    \label{eq: uncertainty of x}
    \begin{split}
    &{\Delta\hat{\underbar{x}}} \simeq \mathcal{S}(I_f \otimes {\underbar{e}})\mathcal{Q} = \mathcal{S}\underbar{e}_{I_f}\mathcal{Q},
    \end{split}
\end{equation}
where $I_f$ is the $f \times f$ identity matrix, and $\underbar{e}_{I_f} = I_f \otimes {\underbar{e}}$.
\\By inserting expressions (\ref{eq: uncertainty on e}) and (\ref{eq: uncertainty of x}) into (\ref{eq: uncertainty on state equation}), the final expression for the uncertainty of the estimates of $A$, $B$ and $K$ in vectorized form is: 
\begin{equation}
\label{eq: uncertainty of A,B,K}
     \begin{split}
    &\operatorname{vec}([\Delta\widehat{A}, \; \Delta\widehat{B}, \; \Delta\widehat{K}]) = \\& = [ \mathcal{S} \underbar{e}_{I_f} \mathcal{Q}\mathcal{\tilde{D}} - \widehat{A}\mathcal{S}\underbar{e}_{I_f}\mathcal{Q} + \widehat{K} \underbar{e}\mathcal{Z}_{ext}^{\dagger}\mathcal{Z}_{ext}]\begin{bmatrix} \hat{\underbar{x}} \\ \underbar{u} \\ \hat{\underbar{e}}  \end{bmatrix}^{\dagger} = \\ &= \left[ \left( \mathcal{Q}\mathcal{\tilde{D}}\begin{bmatrix} \hat{\underbar{x}} \\ \underbar{u} \\ \hat{\underbar{e}}\end{bmatrix}^{\dagger}\right)^\top\otimes \mathcal{S} - \left( \mathcal{Q}\begin{bmatrix} \hat{\underbar{x}} \\ \underbar{u} \\ \hat{\underbar{e}}\end{bmatrix}^{\dagger}\right)^\top\otimes  \widehat{A}\mathcal{S}\right] \operatorname{vec}(\underbar{e}_{I_f}) \\ &+ \left[ \left( \mathcal{Z}_{ext}^{\dagger}\mathcal{Z}_{ext} \begin{bmatrix} \hat{\underbar{x}} \\ \underbar{u} \\ \hat{\underbar{e}}\end{bmatrix}^{\dagger} \right)^\top \otimes \widehat{K}  \right] \operatorname{vec}(\underbar{e}) = \\ & = \alpha_1\operatorname{vec}(\underbar{e}_{I_f}) + \beta_1\operatorname{vec}(\underbar{e}) \, ,
    \end{split}
\end{equation}
where the identity $\operatorname{vec}(\mathcal{A}\mathcal{B}\mathcal{C}) = (\mathcal{C}^\top \otimes \mathcal{A}) \operatorname{vec}(\mathcal{B})$, with $\mathcal{A}$, $\mathcal{B}$, $\mathcal{C}$ generic matrices, is used.
Analogously, the uncertainty of the estimates of $C$ and $D$ is:
\begin{equation}
\label{eq: uncertainty of C,D}
    \begin{split}
        &\operatorname{vec}([\Delta\widehat{C} \; , \; \Delta\widehat{D}]) = [-\widehat{C}\mathcal{S}\underbar{e}_{I_f}\mathcal{Q} + \underbar{e}\mathcal{Z}_{ext}^{\dagger}\mathcal{Z}_{ext}] \begin{bmatrix} \hat{\underbar{x}} \\ \underbar{u} \end{bmatrix}^{\dagger} = \\ &= \left[\left(\mathcal{Q}\begin{bmatrix} \hat{\underbar{x}} \\ \underbar{u} \end{bmatrix}^{\dagger} \right)^\top \otimes (-\widehat{C}\mathcal{S})\right]\operatorname{vec}(\underbar{e}_{I_f}) + \\& + \left[\left(\mathcal{Z}_{ext}^{\dagger}\mathcal{Z}_{ext}\begin{bmatrix} \hat{\underbar{x}} \\ \underbar{u} \end{bmatrix}^{\dagger} \right)^\top \otimes I \right]\operatorname{vec}(\underbar{e}) = \\& = \alpha_2\operatorname{vec}(\underbar{e}_{I_f}) + \beta_2 \operatorname{vec}(\underbar{e}) \; .
    \end{split}
\end{equation}
\subsection{Asymptotic variance expressions}
To derive the expressions of the asymptotic variance from equations (\ref{eq: uncertainty of A,B,K}) and (\ref{eq: uncertainty of C,D}), an operator $\mathcal{P}$ that relates $\operatorname{vec}(\underbar{e})$ to $\operatorname{vec}(\underbar{e}_{I_f})$, such that $\operatorname{vec}(\underbar{e}_{I_f}) = \mathcal{P}\operatorname{vec}(\underbar{e})$, has to be introduced (see \cite{VANWINGERDEN2012ASYMPTOTICVARIANCE} for further details). This operator acts on $\alpha_1$ and $\alpha_2$ as follows:
\begin{equation}
\begin{split}
\alpha_1 P &=
\sum_{i=1}^{f}
\left(
\mathcal{Q}_i
\mathcal{\tilde{D}}\begin{bmatrix} \hat{\underbar{x}} \\ \underbar{u} \\ \hat{\underbar{e}}\end{bmatrix}^{\dagger}\right)^\top
\otimes \mathcal{S}_i - \left( \mathcal{Q}_i\begin{bmatrix} \hat{\underbar{x}} \\ \underbar{u} \\ \hat{\underbar{e}}\end{bmatrix}^{\dagger}\right)^\top\otimes  \widehat{A}\mathcal{S}_i \;,
\\
\alpha_2 P &=
\sum_{i=1}^{f}
\left(\mathcal{Q}_i\begin{bmatrix} \hat{\underbar{x}} \\ \underbar{u} \end{bmatrix}^{\dagger} \right)^\top \otimes (-\widehat{C}\mathcal{S}_i)\;,
\end{split}
\end{equation}
where $[\mathcal{S}_1 \; \; \mathcal{S}_2 \; \cdots \; \mathcal{S}_f ] = \mathcal{S}$ and $[\mathcal{Q}_1^\top \; \; \mathcal{Q}_2^\top \; \cdots \; \mathcal{Q}_f^\top ] = \mathcal{Q}^\top$, with $\mathcal{S}_i \in \mathbb{R}^{n_x \times n_y}$ and $\mathcal{Q}_i \in \mathbb{R}^{n_c \times n_c}$. Hence,  (\ref{eq: uncertainty of A,B,K}) and (\ref{eq: uncertainty of C,D}) can be rewritten compactly as:
\begin{equation}
        \operatorname{vec}([\Delta\widehat{A}, \; \Delta\widehat{B}, \; \Delta\widehat{K} \; ,\Delta\widehat{C}, \; \Delta\widehat{D}]) = \begin{bmatrix} \alpha_1P + \beta_1 \\ \alpha_2P + \beta_2 \end{bmatrix} \operatorname{vec}(\underbar{e}). 
\end{equation}
Recalling that $\Delta\widehat{\Theta} =  \operatorname{vec}([\Delta\widehat{A}, \; \Delta\widehat{B}, \; \Delta\widehat{K}, \; \Delta\widehat{C}, \; \Delta\widehat{D}])$, the asymptotic variance of the state-space matrices is given by:
\begin{equation}
\label{eq: asymptotic variance}
    \begin{split}
        &\operatorname{plim}\Delta\widehat{\Theta}\Delta\widehat{\Theta}^\top = \\ &= \begin{bmatrix} \alpha_1P + \beta_1 \\ \alpha_2P + \beta_2 \end{bmatrix}\mathbb{E}[ \operatorname{vec}(\underbar{e})\operatorname{vec}(\underbar{e})^\top]\begin{bmatrix} \alpha_1P + \beta_1 \\ \alpha_2P + \beta_2 \end{bmatrix}^\top \\ &= \begin{bmatrix} \alpha_1P + \beta_1 \\ \alpha_2P + \beta_2 \end{bmatrix}(I \otimes \Lambda)\begin{bmatrix} \alpha_1P + \beta_1 \\ \alpha_2P + \beta_2 \end{bmatrix}^\top,
    \end{split}
\end{equation}
where the last expression follows from the sequence $\underbar{e}$ being white, which implies that its covariance matrix can be computed as $\Lambda = \hat{\underbar{e}} \hat{\underbar{e}}^\top/n_c\in\mathbb{R}^{n_y\times n_y}$.
\section{Hermite-domain validation}
\label{sec5: Asymptotic variance validation in the Hermite domain}

Equations (\ref{eq: asymptotic variance}) are first validated in the Hermite domain, so as to exclude projection errors due to finite-dimensional Hermite expansions. This setting also allows one to impose a white innovation sequence and avoid numerically induced coloring, consistently with the white-noise innovation property shown in the Appendix. Moreover, the true state-space matrices are used in (\ref{eq: asymptotic variance}) to remove identification bias.

\subsection{Hermite-domain simulation}
To address the issue of simulating in the Hermite domain, the state equation is written as:
\begin{equation}
    \label{eq: LTI dynamics in output-error form}
    \underbar{x} \mathcal{D}' = \frac{A + \beta I}{\gamma} \underbar{x} + \frac{B}{\gamma} \underbar{u} + \frac{K}{\gamma}\underbar{e}
    = \tilde{A}\underbar{x} + \tilde{B}\underbar{u} + \tilde{K}\underbar{e}.
\end{equation}
This is a Sylvester-type algebraic matrix equation ($\mathcal{X}\mathcal{A}+\mathcal{B}\mathcal{X}=\mathcal{C}$) in the unknown $\underbar{x}$. Using vectorization,
\begin{equation}
\label{eq:vec identities hermite simulation}
\begin{split}
\mathrm{vec}(\underbar{x}\mathcal{D}') &= (\mathcal{D}'^\top \otimes I_{n_x})\,\mathrm{vec}(\underbar{x}), \\
\mathrm{vec}(\tilde{A}\underbar{x}) &= (I_{n_c}\otimes \tilde{A})\,\mathrm{vec}(\underbar{x}),
\end{split}
\end{equation}
so that:
\begin{equation}
\label{eq:hermite_linear_system}
\left(\mathcal{D}'^\top \otimes I_{n_x} - I_{n_c}\otimes \tilde{A}\right)\mathrm{vec}(\underbar{x})
=
\mathrm{vec}(\tilde{B}\underbar{u}+\tilde{K}\underbar{e}).
\end{equation}
By defining:
\begin{equation}
\label{eq:M and b hermite simulation}
\begin{split}
M = \mathcal{D}'^\top \otimes I_{n_x} - I_{n_c}\otimes \tilde{A},
\qquad
b = \mathrm{vec}(\tilde{B}\underbar{u}+\tilde{K}\underbar{e}),
\end{split}
\end{equation}
one obtains the linear system:
\begin{equation}
\label{eq:linear system hermite simulation}
M\,\mathrm{vec}(\underbar{x}) = b.
\end{equation}
Solving (\ref{eq:linear system hermite simulation}) and reshaping $\mathrm{vec}(\underbar{x})$ yields the Hermite-domain state trajectory $\underbar{x}$. The corresponding output is then generated as:
\begin{equation}
\label{eq: noise-free dynamics}
\tilde{\underbar{y}} = C\underbar{x} + D\underbar{u},
\end{equation}
to which a white Gaussian innovation $\underbar{e}$ is added.

\subsection{Numerical validation}
\label{sec: Numerical validation of Hermite simulation}

\begin{figure}[htbp]
	\begin{center}
		\includegraphics[width=\columnwidth]{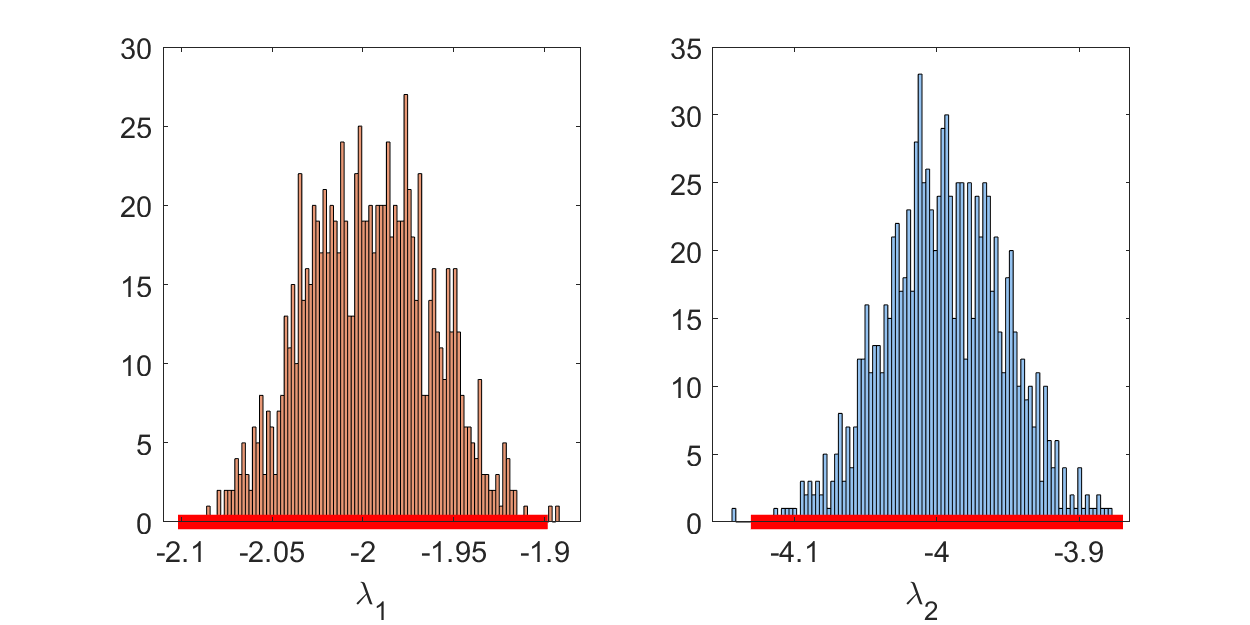}
		\caption{Comparison between the analytical three-standard-deviation bounds (red) and the eigenvalues obtained from 1000 MC runs.}
		\label{fig:hysto Hermite}
	\end{center}
\end{figure}

The validation is performed on the LTI system
\begin{equation}
\begin{split}
\label{eq: example system matrices}
&A = \begin{bmatrix} -2 & 1 \\ 0 & -4 \end{bmatrix}, \;
B = \begin{bmatrix} 1 \\ 1 \end{bmatrix}, \;
K = \begin{bmatrix} 0.0119 & -0.0143 \\ -0.0238 & 0.0119 \end{bmatrix}, \\
&C = \begin{bmatrix} 1 & 0 \\ 4 & 1 \end{bmatrix}, \;
D = \begin{bmatrix} 0 \\ 0 \end{bmatrix},
\end{split}
\end{equation}
whose true eigenvalues are $\lambda_1=-2$ and $\lambda_2=-4$.
\\The input $\underbar{u}$ is obtained by projecting onto $\{h_n\}_{n=0}^{n_{max}}$, with $n_{max}=250$, the time-domain linear sine sweep:
\begin{equation}
\label{eq:linear sine sweep}
u(t)=\sin\!\left[2\pi\left(f_1 t+\frac{f_2-f_1}{2T}t^2\right)\right],
\end{equation}
with $T=10\,\mathrm{s}$, $f_1=0\,\mathrm{Hz}$, and $f_2=8\,\mathrm{Hz}$. The innovation covariance is chosen as $\Lambda = \operatorname{diag}(\sigma_1^2, \sigma_2^2) = \operatorname{diag}(1 \times 10^{-6}, 1 \times 10^{-6})$.
The asymptotic variance predicted by (\ref{eq: asymptotic variance}), evaluated using the true matrices and the known covariance $\Lambda$, is converted into three-standard-deviation bounds on the system eigenvalues. These bounds are compared with a MC simulation of 1000 runs, each obtained with a different realization of $\underbar{e}$. For each pair $(\underbar{u},\underbar{y})$, the HD-PBSID algorithm is applied and the eigenvalues of $\widehat{A}$ are computed. As shown in Fig.~\ref{fig:hysto Hermite}, the analytical bounds closely match the MC eigenvalue distribution.

\begin{rmk}\label{rmk:remark1}
Using (\ref{eq: expression of delta CK^p}), the state-sequence uncertainty can also be written as ${\Delta\hat{\underbar{x}}} \simeq \Delta\widehat{\mathcal{K}^p}\mathcal{Z} = C^{-1}\underbar{e} Z_{\mathrm{ext,part}}^{\dagger}$.
This expression is used here in place of (\ref{eq: uncertainty of x}) to avoid bias in the validation setting.
\end{rmk}
\section{Time-domain validation}
\label{sec6: Test example}
The asymptotic variance expression in (\ref{eq: asymptotic variance}) is now assessed in a more realistic setting using noisy time-domain data. In this case, the estimates may be affected by bias, projection errors, and possible coloring of the innovation sequence. In particular, the white-noise condition on the innovation sequence underlying the derivation is no longer guaranteed to hold exactly.
\\Two treatments of the innovation sequence $\underbar{e}$ are considered. In the first one, $\underbar{e}$ is estimated directly from the residuals through (\ref{eq:innovation data matrix estimation}), and its covariance is approximated as
$\Lambda=(\hat{\underbar{e}} - \hat{\overline{\underbar{e}}})(\hat{\underbar{e}} - \hat{\overline{\underbar{e}}})^\top/n_{max}$, with $\hat{\overline{\underbar{e}}}$ being the column-wise sample mean of $\hat{\underbar{e}}$.
This covariance is then used in (\ref{eq: asymptotic variance}). Since the residual-based innovation is not guaranteed to be white, the white-noise condition may be violated. In the second treatment, the same covariance $\Lambda$ is retained, but a white innovation sequence is generated from it. This re-sampled innovation is then used to recompute $\widehat{A}$, $\widehat{B}$, and $\widehat{K}$ through (\ref{eq:final least squares}), together with all the quantities entering (\ref{eq: asymptotic variance}). This second treatment explicitly enforces the innovation sequence to be white and generally yields more realistic estimates of the Kalman gain, leading to more reliable uncertainty bounds.
\\As a first test case, the matrices $A$, $B$, $C$, and $D$ are chosen as in (\ref{eq: example system matrices}). When the Signal-to-Noise Ratio SNR $=19.54$ dB, the HD-PBSID algorithm identifies the eigenvalues $\hat{\lambda}_1=-1.97$ and $\hat{\lambda}_2=-3.98$, showing a small bias with respect to the true poles. The identified matrices, the estimated innovation sequence, and the estimated state sequence from (\ref{eq:state sequence estimate via SVD}) are inserted into (\ref{eq: asymptotic variance}) to compute analytical uncertainty bounds, expressed as three standard deviations around the identified eigenvalues.
These bounds are compared with the empirical eigenvalue distributions obtained from 1000 MC runs with varying noise realizations. The comparison is shown in Fig.~\ref{fig:hysto numerical SNR 19.54, residual innovation}. When the innovation is taken directly from the residuals, the analytical bounds overestimate the empirical spread of the identified eigenvalues. This is accompanied by a large estimated Kalman gain, 
$\widehat{K} =
[
-13.6266 , -5.3800 ;
19.5279 , -43.2314]
$,
which inflates the predicted uncertainty.
When the innovation is instead enforced to be white, the estimated Kalman gain becomes 
$\widehat{K} =
[
0.3199  ,  1.1065 ;
   -3.1766  , -2.7810
]$,
which is more consistent with the expected system behavior. As a result, the predicted asymptotic variance decreases and the corresponding bounds become significantly tighter, in much better agreement with the MC results, as shown in Fig. \ref{fig:hysto numerical SNR 19.54, white-noise innovation}.
\begin{figure}[htbp]
    \centering
    \begin{subfigure}[b]{\columnwidth}
    \begin{center}
		\includegraphics[width=\columnwidth]{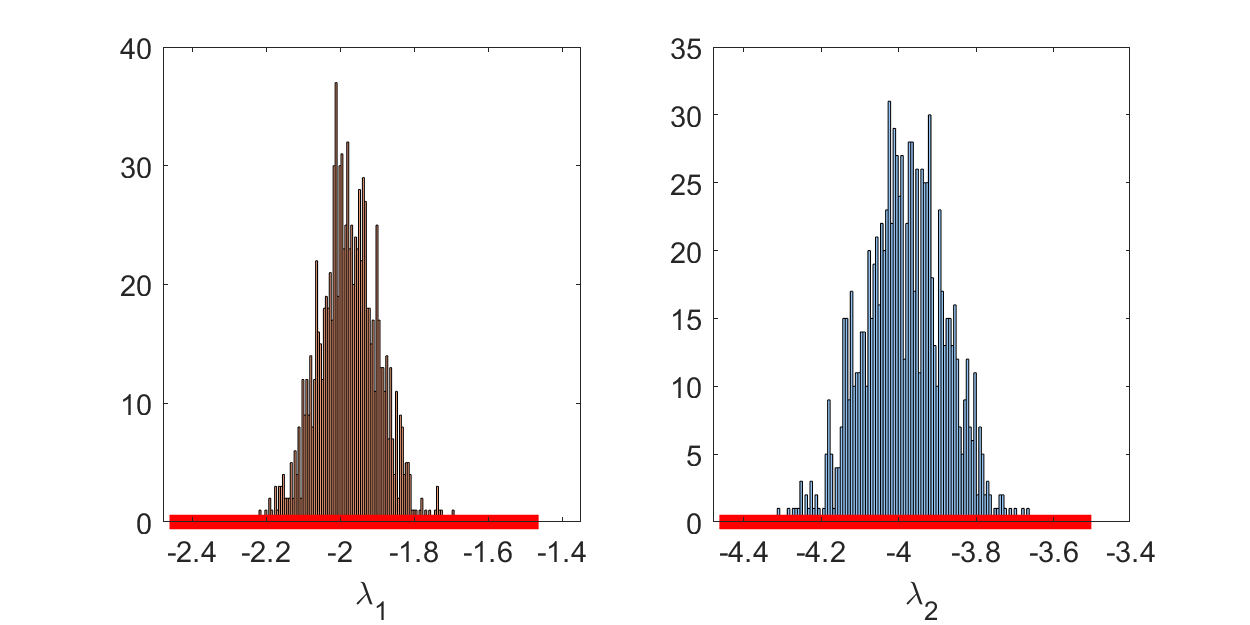}
		\caption{}
		\label{fig:hysto numerical SNR 19.54, residual innovation}
	\end{center}
    \end{subfigure}
    \begin{subfigure}[b]{\columnwidth}
    \begin{center}
		\includegraphics[width=\columnwidth]{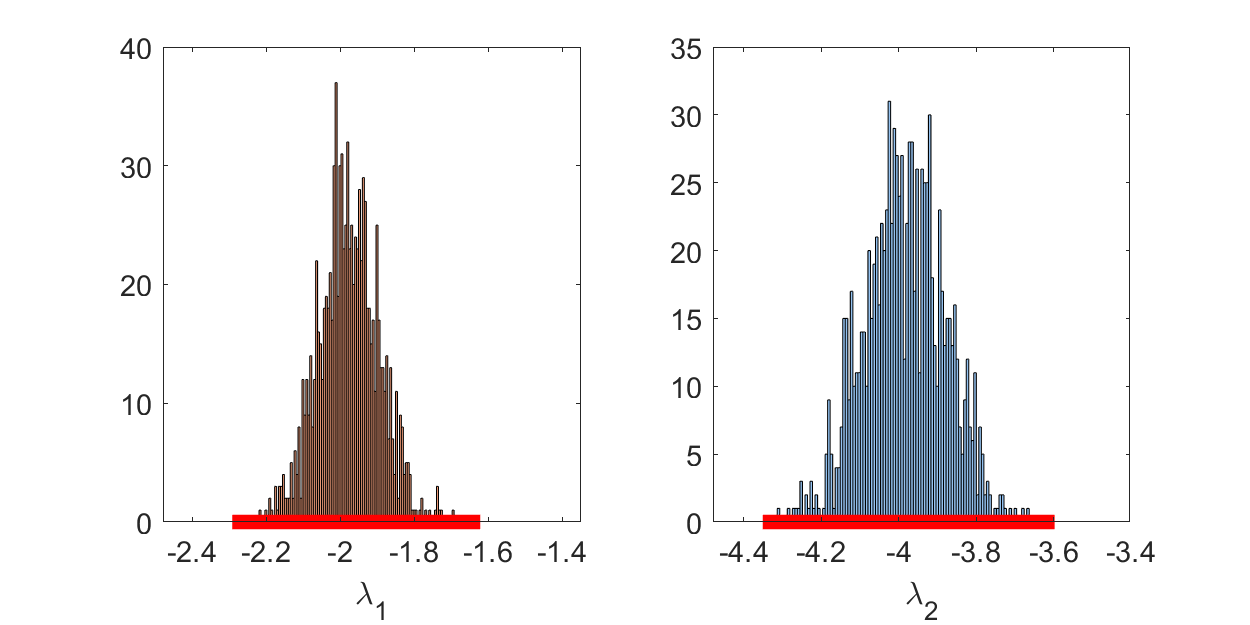}
		\caption{}
		\label{fig:hysto numerical SNR 19.54, white-noise innovation}
	\end{center}
    \end{subfigure}
   \caption{Comparison between the analytical three-standard-deviation bounds (in red) and the eigenvalues obtained via MC simulation, for (a) residual-based innovation and (b) enforced white innovation.}
\end{figure}
A broader quantitative comparison is reported in Table~\ref{tab:time_std_comparison}, which compares analytical and MC standard deviations across SNR levels and pole locations for both innovation treatments.
In all cases, the residual-based innovation yields conservative estimates and tends to overestimate the empirical variability of the identified eigenvalues. When the innovation is enforced to be white, the discrepancy is substantially reduced. These results indicate that the proposed asymptotic variance expressions remain informative also in the time-domain setting, but their accuracy seems to depend strongly on how closely 
the innovation sequence satisfies the whiteness condition.
\begin{table}[t]
\caption{Analytical and MC three standard deviations ($3\sigma$) of the identified eigenvalues for different SNR levels and pole locations. R and W denote residual-based and (enforced) white innovation, respectively.}
\label{tab:time_std_comparison}
\centering
\begin{tabular}{c c c c c }
\hline
Innov. & SNR & Poles & $3\sigma_{\mathrm{analyt}}$ & $3\sigma_{\mathrm{MC}}$  \\
\hline
R & 29.54 & $(-2,\,-4)$   & $(0.167 ,0.169)$ & $(0.075,0.105)$ \\
W  & 29.54 & $(-2,\,-4)$     & $(0.105, 0.129)$ & $(0.075,0.105)$  \\
R & 19.54 & $(-2,\,-4)$     & $(0.514,0.490)$ & $(0.242,0.312)$  \\
W  & 19.54 & $(-2,\,-4)$     & $(0.335,0.377)$ & $(0.242 ,0.312)$ \\
R & 31.89 & $(-0.1,\,-10)$  & $(0.101,0.151)$ & $(0.016,0.142)$  \\
W  & 31.89 & $(-0.1,\,-10)$  & $(0.017,0.138)$ & $(0.016,0.142)$ \\
R & 21.86 & $(-0.1,\,-10)$  & $(0.119, 0.393)$ & $(0.019, 0.362)$  \\
W  & 21.86 & $(-0.1,\,-10)$  & $(0.022,0.373)$ & $(0.019, 0.362)$  \\
\hline
\end{tabular}
\end{table}

\section{CONCLUSIONS}
\label{sec7: Conclusions}
This paper derives analytical expressions for the uncertainty and asymptotic variance of the CT state-space matrices identified by HD-PBSID. In an ideal validation setting, where bias and the coloring of the innovation are excluded, the resulting eigenvalue uncertainty bounds closely match the MC eigenvalue distributions, supporting the validity of the obtained formulas. In a more realistic numerical example, the analytical bounds are found to overestimate the observed spread of the identified eigenvalues when the white-noise innovation condition is only approximately satisfied. However, when the innovation is forced to be a white noise process, the predictions on the eigenvalue distributions significantly improve, which suggests that the obtained asymptotic variance formulas can also be effectively used on time-domain data and in the presence of a small amount of bias. 


\addtolength{\textheight}{-12cm}   



\section*{APPENDIX}
\label{appendix}

Let $de(t)\in\mathbb{R}$ be a continuous-time white-noise process on $\mathbb{T} = [-\mathcal{T},\mathcal{T}]$ such that
$\mathbb{E}[de(t)\,de(s)] = \sigma^2 \delta(t-s)\,dt\,ds$,
and let $\{h_n\}$ be an orthonormal Hermite basis on the same interval, i.e.
$
\int_{-\mathcal{T}}^{\mathcal{T}} h_n(t)h_m(t)\,dt=\delta_{nm}$.
Define the Hermite coefficients as
$e_n=\int_{-\mathcal{T}}^{\mathcal{T}} h_n(t)\,de(t)$.
Their covariance is
\begin{equation}
\begin{split}
\mathbb{E}[e_n e_m]
&=
\int_{-\mathcal{T}}^{\mathcal{T}}\!\!\int_{-\mathcal{T}}^{\mathcal{T}}
h_n(t)h_m(s)\,\mathbb{E}[de(t)\,de(s)] \\
&=
\sigma^2
\int_{-\mathcal{T}}^{\mathcal{T}}\!\!\int_{-\mathcal{T}}^{\mathcal{T}}
h_n(t)h_m(s)\,\delta(t-s)\,dt\,ds \\
&=
\sigma^2 \int_{-\mathcal{T}}^{\mathcal{T}} h_n(t)h_m(t)\,dt
=
\sigma^2 \delta_{nm}.
\end{split}
\end{equation}
Therefore, the Hermite-domain innovation coefficients form a white sequence with variance $\sigma^2$. The multidimensional case follows analogously.


\bibliographystyle{IEEEtran}
\bibliography{bibliography}

\end{document}